\documentstyle[preprint,aps]{revtex}
\begin{document}
%\draft
%\twocolumn
\title{Laying the ghost of the relativistic temperature transformation}
\author{Peter T.\ Landsberg\thanks{
                                   Permanent address: Faculty of Mathematical 
				   Studies, University of Southampton, 
				   Southampton SO17 1BJ, England.} 
			   \thanks{ e-mail: ptl@maths.soton.ac.uk}  
       and George E.A.\ Matsas\thanks{ e-mail: matsas@axp.ift.unesp.br}
      }
\address{Instituto de F\'\i sica Te\'orica, Universidade Estadual Paulista\\ 
         Rua Pamplona 145, 01405-900--S\~ao Paulo, SP\\ 
	 Brazil}

\maketitle

%%%%%%%%%%%%%%%%%%%%%%%%%%%%%%%%%%%%%%%%%%%%%%%%%%%%%%%%%%%%%%%%%%%%%%%%%%%%%
%%%%%%%%%%%%%%%%%%%%%%%%%%%%    Abstract    %%%%%%%%%%%%%%%%%%%%%%%%%%%%%%%%%
%%%%%%%%%%%%%%%%%%%%%%%%%%%%%%%%%%%%%%%%%%%%%%%%%%%%%%%%%%%%%%%%%%%%%%%%%%%%%

\begin{abstract}
Using the Unruh-DeWitt detector, it is shown that a universal and
continuous Lorentz transformation of temperature cannot exist
for black-body radiation. Since any valid Lorentz transformation
of temperature must be able to deal with black-body radiation,
it is concluded that a universal and continuous temperature
transformation does not exist.

\end{abstract}

%%%%%%%%%%%%%%%%%%%%%%%%%%%%%%%%%%%%%%%%%%%%%%%%%%%%%%%%%%%%%%%%%%%%%%
%%%%%%%%%%%%%%%%%%%%%%%%%%%%%%%%%%%%%%%%%%%%%%%%%%%%%%%%%%%%%%%%%%%%%%
%%%%%%%%%%%%%%%%%%%%%%%%%%%%%%%%%%%%%%%%%%%%%%%%%%%%%%%%%%%%%%%%%%%%%%
\newpage
\narrowtext
This note represents yet another attempt to lay the ghost of
the relativistic temperature transformation which has motivated
a number of papers by  the founders of relativity and quantum
mechanics \cite{E}--\cite{VL}, more than 60 contributions during
1963-1968 \cite{A2} (see e.g. \cite{O}--\cite{Ka}), 
and a continuing trickle ever since 
(see e.g. \cite{B}--\cite{FV} and references
therein). The new result which makes the present considerations worthwhile
is the explicit formula for the excitation rate of an
Unruh--DeWitt detector, {\em i.e.} a two-level monopole
\cite{UDW}, which has a single proper energy gap $\hbar
\omega'$, and detects massless scalar particles or,
equivalently, ``spinless photons''. Suppose that black-body
radiation with proper temperature $T$ is at rest in some
inertial frame $S$. The excitation rate of an Unruh-DeWitt  
detector moving with constant velocity $v$ through it is
found, with the aid of quantum field theory, to be 
proportional to the particle number density \cite{CM}
\begin{equation}
n'(\omega', T, v)  d\omega'   =   
                                \frac{ \omega' k T 
				\sqrt{1-v^2/c^2} }{4 \pi^2 c^2 v \hbar } 
			         \ln\left[
                       \frac{1-e^{- 
		                  (\hbar \omega' \sqrt{1+v/c})/ 
			          ( k T \sqrt{1-v/c} ) 
		                 }
		            }
                           {1-e^{- 
			          (\hbar \omega' \sqrt{1-v/c})/
				  ( k T \sqrt{1+v/c} ) 
				 }
		            }
			              \right] d\omega'.
\label{Crucial}
\end{equation}
Here we use ``$\, ' \, $'' to refer to quantities as
measured in $S'$, which is the inertial frame in which the
detector is assumed to be at rest.  This
result, while reducing to the Planckian form
\begin{equation}
n (\omega, T) d\omega = 
\frac{\omega^2/c^3}{2 \pi^2 (e^{\hbar \omega /k T}-1)} d\omega 
\label{Pl}
\end{equation}
in the limit  $v \to 0$, {\em has not itself this form.} 

By a black body we mean a system which has 
a Planckian spectrum, and by black-body temperature the 
parameter which characterizes this spectrum;
so if one looks for the temperature of a black-body
as defined in the moving frame $S'$, one needs to
express (\ref{Crucial}) in the form  
\begin{equation}
n'_{bb} (\omega', T') d\omega' = 
\frac{{\omega'}^2/c^3}{2 \pi^2 (e^{\hbar \omega' /k T'}-1)}
d\omega' , 
\label{sorry}
\end{equation}
for some continuous function  $T'=T'(T,v)$. If one succeeds, 
then one would be able to say that temperature transforms under boosts
according to the law $T'=T'(T,v)$. {\em However,  
such a transformation is easily shown mathematically  
not to exist.}  In particular, the popular transformations 
\begin{equation}
T'(T,v)= \gamma^a T
\label{popular}
\end{equation} 
where $\gamma \equiv (1-v^2/c^2)^{-1/2}$, and 
$a=-1$ (see e.g. \cite{E}-\cite{VL}, \cite{Pat}, \cite{FV}), 
$a=+1$ (see e.g \cite{O}, \cite{K}) or 
$a=0$ (see e.g. \cite{L1}, \cite{Ka}, \cite{L2}) 
are useless in this context because, again, they do not reduce
(\ref{Crucial}) to (\ref{sorry}).  
Since any universal continuous relativistic temperature
transformation has to be able to deal at least 
with the black-body case, we conclude that {\em such a transformation
does not exist.} 

What {\em can} be said  is that  an observer at rest in $S'$, and
looking at the radiation in a frequency interval $d\omega'$
coming from the solid angle interval $d\Omega'$, finds indeed a
variant of (\ref{sorry}):
\begin{equation}
{n}'_{\theta'} (\omega', T'_{\theta'}) d\omega' d\Omega' =  
\frac{{\omega'}^2/c^3 }{2\pi^2 (e^{\hbar \omega' /k T'_{\theta'}}-1)}  
d\omega' d\Omega' .
\label{Pl*}
\end{equation}
Here $T'_{\theta'}$ is a {\em ``directional''}  temperature 
defined by
\begin{equation} 
T'_{\theta'}(T,v,\theta') = \frac{T \sqrt{1-v^2/c^2}}{1- \cos
\theta' \, v/c },
\label{TVTETA}
\end{equation}
and  $\theta'$ is the angle between the axis of motion and the
direction of observation \cite{Pau}. (This result was rediscovered later in
connection with studies of the $2.7\, K$ background radiation, e.g.
\cite{BCPWHFSP}, and so has become well--known.) 
Thus, it should be clear that, regrettably, 
one cannot discover the desired transformation $T'(T,v)$ 
used in Eq. (\ref{sorry}) by averaging the elegant result 
(\ref{Pl*}) over solid angles. One finds instead
\begin{equation}
\frac{1}{4\pi} \int {n}'_{\theta'}(\omega', T'_{\theta'})
d\omega' d\Omega' = n'(\omega',T,v) d\omega'.
\end{equation}
This states what is at first sight a surprising result:
Eq. (\ref{Crucial})  is found exactly; this
time, therefore, without any appeal to quantum field theory.

Any definition based on some operational procedure, 
like: {\em ``Temperature is
what some pre-chosen device in $S'$ measures''} will be
arbitrary because different ``thermometers'', and
measuring procedures will lead to different functional
dependencies. For
example, if we construct a thermometer which measures the
average of (\ref{TVTETA}), $T'_{\rm{aver1}} (T, v) \equiv
\langle T'_{\theta'} \rangle$, which is perfectly possible, 
its reading would satisfy
\begin{equation}
T'_{\rm{aver1}} (T, v) = \frac{1}{4\pi} \int T'_{\theta'}
d\Omega' = \frac{cT}{2v\gamma} \ln\frac{1+v/c}{1-v/c} .
\end{equation}
Another thermometer constructed to measure 
$T'_{\rm{aver2}} (T, v) \equiv
\langle {T'}^4_{\theta'} \rangle^{1/4}$, leads, of course, to a
completely different result (p. 31 of \cite{LPPT})
\begin{equation}
T'_{\rm{aver2}} (T, v) = \left[ 
                              \frac{1}{4\pi} \int {T'}^4_{\theta'}
                              d\Omega' 
		            	 \right]^{1/4}
			       =  T \left[ 
			       \frac{1+v^2/3c^2}{1-v^2/c^2} 
			          \right]^{1/4} .
\end{equation}
Clearly $T'_{\rm{aver1}} (T, v=0) = T'_{\rm{aver2}} (T, v=0) = T$.
In fact, a good thermometer is a device which measures a temperature
$T$ [see Eq. (\ref{Pl})], when it is at rest with the thermal bath.
Other operational procedures to measure temperature based
on Unruh-DeWitt detectors are equally possible, and also give 
different results \cite{CM}. 

Previous writers were often concerned with the manipulation of Lorentz 
transformations of thermodynamical variables such as energy, volume, 
entropy, etc. This approach leads to doubtful results 
[see the discussion of Eq. (\ref{popular})] 
unless the theory is made intrinsically 
covariant (see e.g. \cite{Ka}, \cite{L3} for former, and \cite{IA} for
more recent formulations).  
In this case four-tensors $S^\mu$, $Q^\mu$, etc, replace some of the 
usual thermodynamical variables like entropy, heat, etc, and their 
significance then goes beyond the normal thermodynamics. 
(For example, by writing the entropy four-flux as $S^\mu = S u^\mu$,
where $u^\mu$ is the four-velocity associated with equilibrium states,
the usual entropy is given by projecting $S^\mu$ on $u^\mu$.)
It will
be noted that our  approach has by-passed this problem since 
we are discussing this issue from a microscopic rather than a 
macroscopic point of view, and no explicitly thermodynamical
arguments are needed. 

It might be thought that a rough way of seeing that no
black-body equivalent temperature exists for a radiation
enclosure moving relative to an observer may be by noting simply
that, because of the angle dependence in (\ref{Pl*}), this
equation cannot be associated with a legitimate thermal bath
(which is necessarily isotropic). While this view is correct, we
have in this paper gone beyond this observation by obtaining the
angle-averaged Eq. (\ref{Pl*}) in the form of Eq.
(\ref{Crucial}). This has the merit of linking the above simple
observations with earlier works and making the case against the
temperature transformation much more explicit.
 
In summary, our main conclusion is that because 
the temperature concept of a black body 
is unavoidably associated with the Planckian thermal spectrum, and 
because a bath which is thermal in an
inertial frame $S$ is non-thermal in an inertial frame $S'$,  
which moves with some velocity $v \neq 0$ with respect to  $S$, 
a universal relativistic temperature transformation
$T'=T'(T,v)$ cannot exist.
Thus the proper temperature $T$ alone is left  
as the only temperature of universal significance. This seems to 
complete a story started 90 years ago \cite{E} of how the {\em usual} 
temperature transforms,  and to conclude 
a controversy \cite{O} of 33 years' standing. 

\vskip 0.5 truecm 
%\begin{flushleft}
{\bf{\large Acknowledgements:}}
PL would like to acknowledge Funda\c c\~ao de Amparo \`a Pesquisa
do Estado de S\~ao Paulo, and the
European Union through contracts ERB CH RXCT 92 0007 and CIPA-CT-92-4026,
and GM would like to acknowledge Conselho Nacional de 
Desenvolvimento Cient\'\i fico e Tecnol\'ogico for partial financial 
support.

%\end{flushleft}

\end{document}